\documentclass{emulateapj}
\usepackage{amsmath}
\usepackage{booktabs}
\makeatletter
\renewcommand{\fps@figure}{tp}
\makeatother

\setkeys{Gin}{width=\linewidth,height=0.9\textheight,keepaspectratio=true}

\newcommand\htwo{H$_{\rm 2}$}
\newcommand\Halpha{H$\alpha$}
\newcommand\Hbeta{H$\beta$}
\newcommand\hanii{H$\alpha$+[N~II]}

\newcommand\sbunits{$\rm erg~cm^{-2} s^{-1} sr^{-1}$}
\newcommand\funits{$\rm erg~cm^{-2}~s^{-1}$}
\newcommand\cmq{$\rm cm^{-3}$}
\newcommand\twomic{\rm 2.12 $\mu$m}

\newcounter{ionstage}

\slugcomment{Submitted to the Astronomical Journal}

\shorttitle{Knots in the Helix Nebula}
\shortauthors{O'Dell et al.}

\received{2006 November 13}
\begin{document}

\title{Determination of the Physical Conditions of the Knots in the Helix Nebula from Optical and Infrared Observations\footnotemark[1]}

\footnotetext[1]{
Based on observations with the NASA/ESA Hubble Space Telescope,
obtained at the Space Telescope Science Institute, which is operated by
the Association of Universities for Research in Astronomy, Inc., under
NASA Contract No. NAS 5-26555.}

\author{C. R. O'Dell}
\affil{Department of Physics and Astronomy, Vanderbilt University, Box 1807-B, Nashville, TN 37235}

\author{W. J. Henney}
\affil{Centro de Radioastronom\'{\i}a y Astrof\'{\i}sica, Universidad Nacional Aut\'onoma de M\'exico, Apartado Postal 3-72,
58090 Morelia, Michaoac\'an, M\'exico}

\and 

\author{G. J. Ferland}
\affil{Department of Physics and Astronomy, University of Kentucky, Lexington, KY 40506}

\email{cr.odell@vanderbilt.edu}

\begin{abstract}
We use new Hubble Space Telescope and archived images to clarify the nature of the ubiquitous
knots in the Helix Nebula, which are variously estimated to contain a significant to majority fraction of the 
material ejected by its central star. 

We employ published far infrared spectrophotometry and existing \twomic\ images to establish that
the population distribution of the lowest ro-vibrational states of \htwo\  is close to the distribution of a  gas in local thermodynamic equilibrium (LTE) at $988\pm119$~K\@. In addition, we present calculations that show that the weakness of the \htwo\ 
0-0 S(7) line is not a reason for making the unlikely-to-be true assumption that \htwo\ emission is caused by shock excitation.

We derive a total flux from the nebula in \htwo\ lines and compare this with the power available
from the central star for producing this radiation. We establish that neither  soft X-rays nor 912--1100~\AA\ 
radiation has enough energy to power the \htwo\ radiation, only the stellar extreme ultraviolet radiation shortward of 912~\AA\
does.  Advection of material from the cold regions of the knots produces an extensive zone where both atomic and molecular hydrogen are found, allowing the \htwo\ to directly be heated by Lyman continuum radiation, thus providing a mechanism that will probably explain the excitation temperature and surface brightness of the \twomic\  cusps and tails.

New images of the knot 378-801 in the \htwo\ \twomic\ line reveal that the \twomic\ cusp lies immediately 
inside the ionized atomic gas zone. This property is shared by material in the ``tail' region. The \htwo\ \twomic\ emission of the cusp
confirms previous assumptions, while the tail's property firmly establishes that the ``tail" structure is an ionization bounded radiation shadow behind the optically thick core of the knot. The new \twomic\ image together
with archived Hubble images is used to establish a pattern of decreasing surface brightness and 
increasing size of the knots with increasing stellar distance. Although the contrast against the background is greater in \twomic\  than in the optical lines, the higher resolution and signal of optical
images remains the most powerful technique for searching for knots.

A unique new image of a transitional region of the nebula's inner disk in the HeII 4686 \AA\  line fails
to show any emission from knots that might have been found in the He$^{++}$ core of the nebula. We
also re-examined high signal-to-noise ratio ground-based telescope images of this same inner region
and found no evidence of structures that could be related to knots.

\end{abstract}


\keywords{Planetary Nebulae:individual(Helix Nebula, NGC 7293)}

\section{INTRODUCTION}

The dense knots that populate the closest bright planetary nebula NGC 7293 (the Helix Nebula) must play an 
important role in mass loss from highly evolved intermediate mass stars and therefore in the nature of enrichment of
the interstellar medium (ISM) by these stars. It is likely that similar dense condensations are ubiquitous among the planetary nebulae 
(O'Dell et~al. 2002) as the closest five planetary nebulae show similar or related structures. They are an important component of the mass
lost by their host stars, for the characteristic mass of individual knots has been reported as $\geq 10^{-5}~M_\odot$ (from CO emission, Huggins et al.\@ 2002), $5.6 \times 10^{-5}~M_\odot$ (from the dust optical depth determination by Meaburn et~al. (1992), adjusted for the improved distance), and about $3.8 \times 10^{-5}$ M$_\odot$ (O'Dell \&\ Burkert 1997, again from the dust optical depth but with better spatial resolution), and their number has been 
variously estimated to be from 3500 (O'Dell \&\ Handron 1996) from optical observations to much larger numbers (23,000 Meixner et~al. 2005, henceforth MX05; 20,000--40,000  
Hora et~al. 2006, henceforth H06) from infrared imaging. Therefore, these condensations contain a significant fraction to a majority
of all the material ejected.  It is an extremely important point to understand if the ISM is being seeded by these 
knots and if they survive long enough to be important in the general properties of the ISM and also the process of formation of new stars. To understand those late phases, long after the knots have escaped the ionizing environment of
their central stars, one must understand their characteristics soon after their formation-which is the subject
of this study.  

There has been a burst of interest in the Helix Nebula and its knots beginning with the lower resolution groundbased
study of Meaburn et~al. (1992) and the Hubble Space Telescope (HST) images at better than 0.1\arcsec\ resolution (O'Dell \&\ Handron 1996, O'Dell \&\ Burkert 1997) in the optical window.  The entire nebula has been imaged in the
\htwo\ {\it v}=1-0 S(1)
2.12 $\mu$m  line at scales and resolutions of about 4\arcsec\ (Speck et~al. 2002), and 1.7\arcsec/pixel (H06), while Huggins
et~al. 2002) have studied one small region at 1.2\arcsec\ resolution, and the NIC3 detector of the NICMOS instrument of
the HST has been used by Meixner et~al. (2004, MX05) to sample several outer regions at about 0.2\arcsec\ resolution.
A lower resolution (~2\arcsec) study in the longer wavelength 0-0 rovibrational lines has imaged the entire 
nebula with the Spitzer Space Telescope (H06), extending a similar investigation by Cox et~al. (1998, henceforth Cox98)
at 6\arcsec/pixel with the Infrared Space Observatory. Radio observations of the CO (Huggins et~al. 2002, Young et~al. 1999) and H~I (Rodr\'{\i}guez et~al. 2002) emission have even lower spatial resolution, but, the high spectral resolution allows one to see emission from individual knots.

The three dimensional model for the Helix Nebula has also evolved during this time. We now know that the inner 
part of the nebula is a thick disk of 500\arcsec\ diameter seen at an angle of about 23\arcdeg\ from the plane of the 
sky (O'Dell et~al. 2004, henceforth OMM04). This disk has a central core of high ionization material traced by He~II emission (4686 \AA), and
a series of progressively lower ionization zones until its ionization front is reached. The more easily visible lower ionization portions of the inner-disk form the inner-ring of the nebula. There are polar plumes of material
perpendicular to this inner disk extending out to at least 940\arcsec\ (OMM04) to both the northwest and southeast. There is an apparent
irregular outer-ring which Meaburn et~al. (2005, henceforth M05) argue is a thin layer of material on the surface of the perpendicular plumes,
whereas OMM04) and O'Dell (2005) argue that this is due to a larger ring lying almost perpendicular 
to the inner disk.  

The nature of the knots has attracted considerable attention. O'Dell \&\ Burkert (1997) determined the properties 
using HST WFPC2 emission line images in \Halpha, [N~II], and [O~III], while O'Dell et~al. (2000, henceforth OHB00) analyzed HST slitless spectra of the bright
knot 378-801 in \Halpha\ and [N~II], an investigation extended in a study (O'Dell et~al, henceforth OHF05) with better slitless
images in the same lines and also the [O~I] line at 6300 \AA. We will adopt the position based designation system
described in O'Dell \&\ Burkert (1997) and the trigonometric parallax distance of 219 pc from Harris et~al.
(2007). 
The object 378-801 is the best studied of the knots and the primary target for the program reported upon 
in this paper.  At 219 pc distance from the Sun, the 1.5\arcsec\ chord of the bright cusp surrounding the neutral central core
of 378-801 is $4.9 \times 10^{15}$ cm. O'Dell \&\ Burkert (1997) estimate that the peak density in the ionized cusp is about
1200 \cmq\ and the central density of the core, derived from the optical depth in dust, is $4.8 \times 10^{5}$ \cmq, a number similar to 
the \htwo\ density of $\geq$10$^{5}$ \cmq\ necessary to produce the thermalized population distribution found for the J states within
the $v=0$ levels of the electronic (X~$\rm ^{1}\Sigma^{+}_{g}$) ground state by Cox98.  Cox98 determined that two sample regions of knots were close to a
population distribution of about 900 K, a similar result is found by an analysis (\S\ 4.2) of new observations (H06) of different
regions of knots.

As was argued in O'Dell \&\ Handron (1996), the knots are neutral condensations ionized on the side facing the central star.
L\'opez-Mart\'{\i}n et~al. (2001) have shown that the early apparent discrepancy between the observed and predicted surface brightness of the bright cusps is resolved once one considers the dynamic nature of the flow from the cusp ionization front, which depresses the recombination emission from the ionized gas.
 The central cores are molecular, being visible in CO (Huggins et~al. 2002), and producing the 
multiple velocity components one sees in the low spatial resolution-high velocity resolution CO studies (Young et~al. 1999).  The 
\htwo\ emission is produced in a thin layer of material immediately behind the ionized cusp (Huggins et~al. 2002).
OHB00 show that the optical structure within the ionized cusp can only be explained if the material is heated
on a timescale that is longer than the time for cool material to flow from the ionization front across the width of 
the cusp. This slow heating rate means that forbidden lines are seen only further away from the ionization front 
(because these require energetic electrons to cause their collisional excitation) whereas recombination lines like
\Halpha\ arise preferentially from the cool regions closer to the ionization front. A succession of papers (Burkert \&\ O'Dell 1998; OHB00, OHF05 ) attempting
to simultaneously account for both the ionization and flow of material have produced a general model where cool
material in the molecular central core flows towards the ionization front, is slowly heated, and upon passing through
the ionization front is more rapidly heated and accelerated. At first examination 
the most recent models look satisfactory. However, the models fail because the theoretically expected zone of 900 K gas 
has much too low a column density 
to account for the observed surface brightness in \htwo\ (OHF05). These \htwo\ observations must be telling us about
something overlooked in previous models.  As we discuss in \S\ 4.4, this missing ingredient seems to be that the earlier models were for a static structure, whereas the knots are actually in a state of active flow.

This paper reports on work intended to clarify the nature of the knots in the Helix Nebula. New observations with all
three of the imaging instruments on the HST were made and are described in \S\ 2, then analyzed in \S\ 3. A new
and more refined theoretical model is presented in \S\ 4, while these and other recent observations and models are discussed 
in \S\ 5.


\section{OBSERVATIONS}
In this investigation we draw on both our own new observations made with the HST and published observations 
made with a variety of telescopes.  These observations range from the optical through the infrared.

\subsection{New Observations}
The new HST observations were made during eight orbits over the period 2006 May 22-24 as program GO 10628 (co-author O'Dell as Principlal Investigator). The Helix Nebula
is sufficiently large that we could simultaneously observe it with the three operating imaging instruments, the WFPC2 
(Holtzman et al.\@ 1995), the NICMOS (Thompson et al.\@ 1998), and the ACS (Gonzaga, S. et al. 2005). By careful selection of 
the pointing and orientation of the spacecraft, we were able to sample three regions that are useful for understanding
the knots and their structure. In each case the new observations are either unique or of substantially longer 
exposure time than previous similar observations. The placement of the fields of view are shown in Figure 1.

\begin{figure*}
  \centering
  \includegraphics{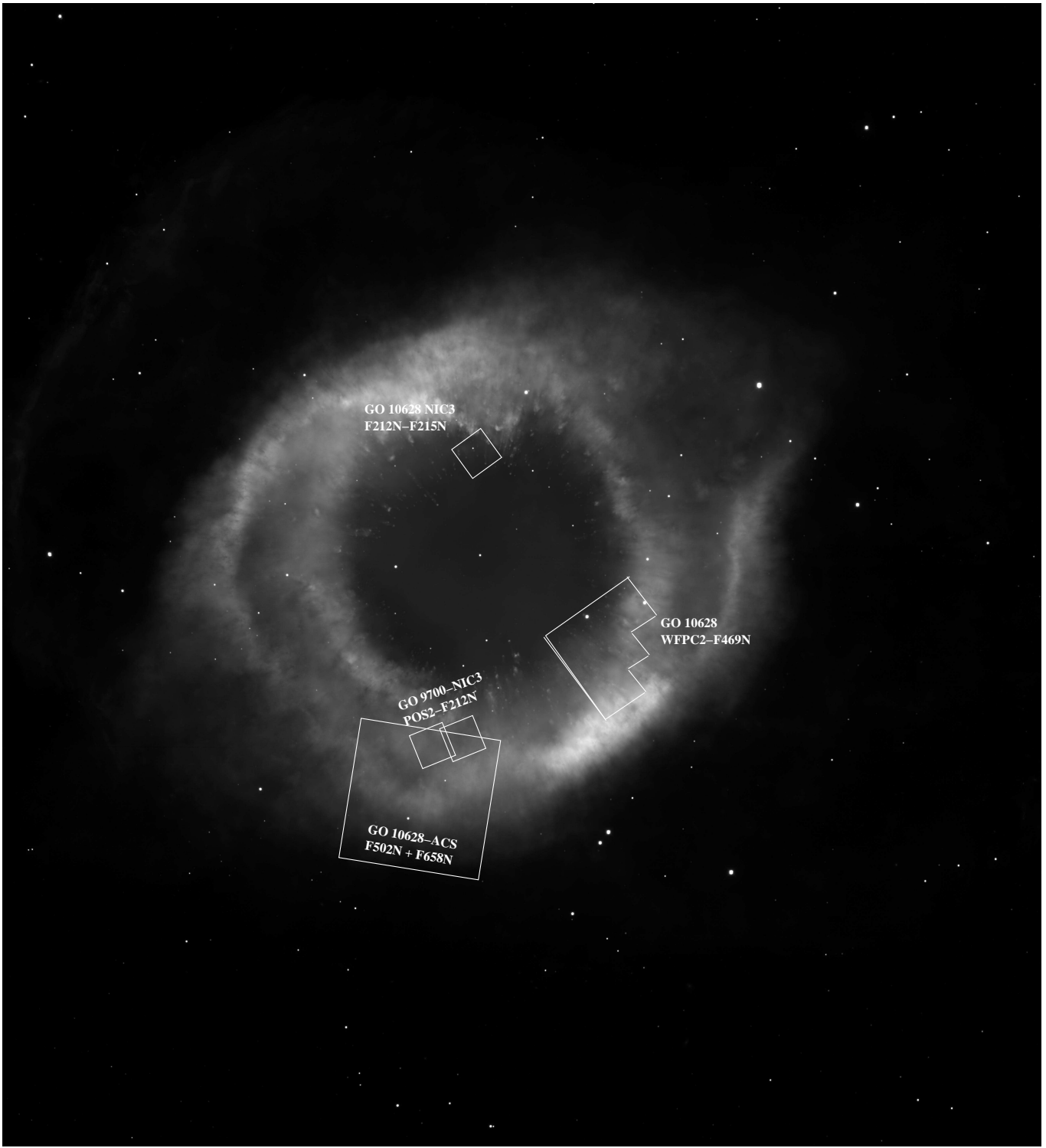}
  \caption{This 1500\arcsec x 1650\arcsec\ image of the Helix Nebula
    in \hanii\ from OMM04 shows the fields covered by the three image
    programs of GO 10628 and also the two NIC3 fields in program GO
    9700 that overlap with the GO 10628 ACS field.}
\end{figure*}

\subsubsection{NICMOS Images in \htwo\ of the Knot 378-801}

The NICMOS observations were made with the NIC3 camera (256x256 pixels, each about 0.2\arcsec /pixel), with eight 
exposures in both the F212N (isolating the 2.12 $\mu$m \htwo\ line and the underlying continuum) and the 
F215N (isolating primarily the underlying continuum as there are no strong lines in this region) filters. Each of the
sixteen exposures was 1280 s duration. The pointing was changed in a four position pattern, with steps of
5\arcsec.  The On-the-Fly standard processing image products were our starting point. These images were combined using
IRAF
\footnote{IRAF is distributed by the National Optical
Astronomy Observatories, which is operated by the Association of
Universities for Research in Astronomy, Inc.\ under cooperative
agreement with the National Science foundation.}
and tasks from the HST data processing STSDAS package provided by the Space Telescope Science Institute (STScI). The method of calibration used in OHF05 was adopted, where the nebular and 
instrumental continuum was subtracted using the signal from the F215N filter.  

The resultant image is shown in Figure 2  along with aligned WFPC2 optical emission line images
from programs GO 5086 and 5311. The knot 378-801 is located in the low central region of each image and details of its bright
cusp and tail are analyzed in \S\ 3.1 and \S\ 3.2. It is notable that all of the \htwo\ cusps have corresponding \Halpha\ and
[N~II] counterparts and that all of the \Halpha\ and [N~II] cusps have \htwo\ counterparts on these well exposed,
high resolution images. 

\begin{figure*}\centering
  \includegraphics{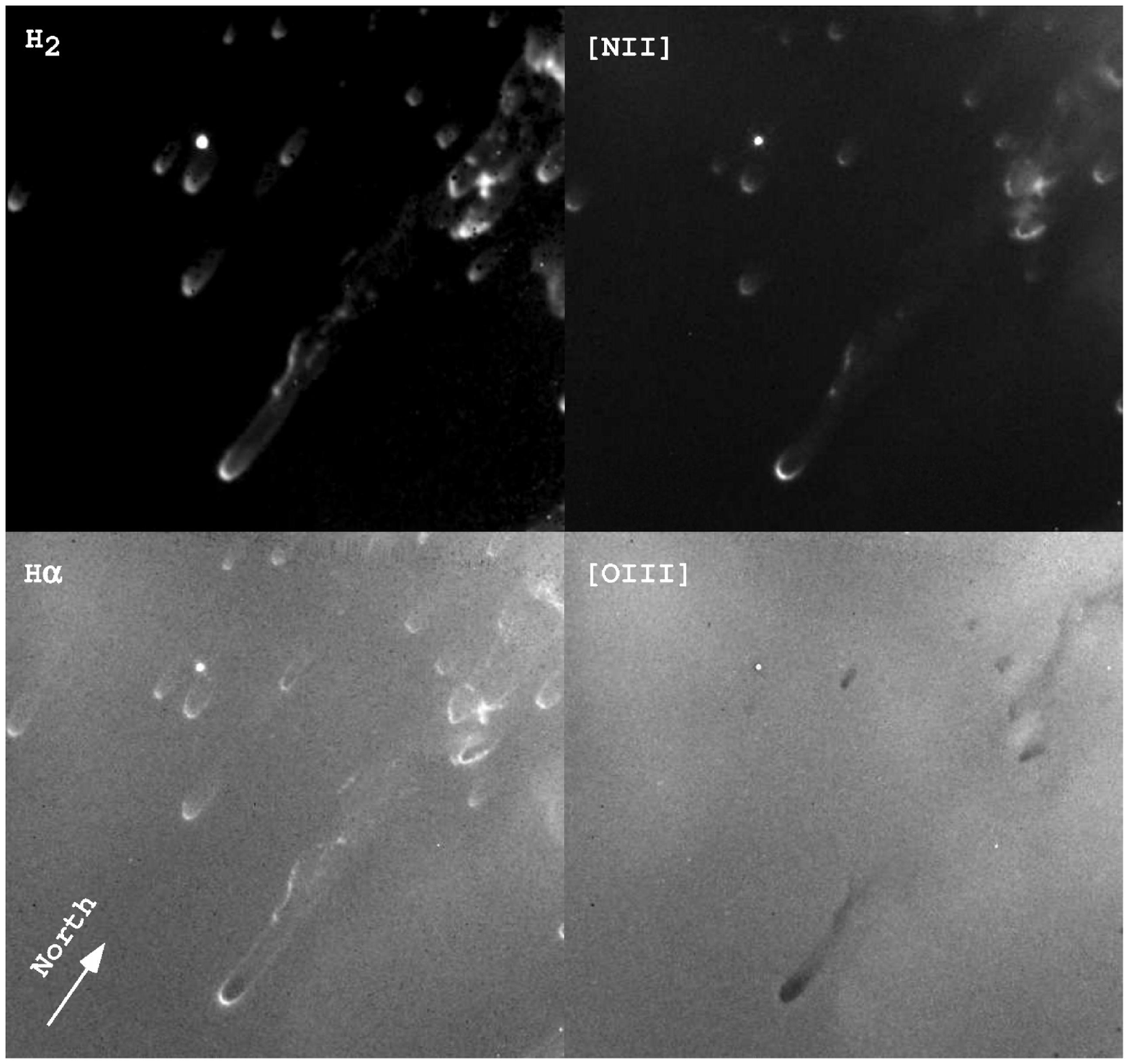}
  \caption{Each image in this mosaic depicts the same 47.5\arcsec x
    44.8\arcsec region targeted with the NICMOS NIC3 camera.  The
    \htwo\ image Jis from GO 10628, as described in the text, while the
    optical emission line images are from WFPC2 programs GO 5086 and
    5311.  The vertical axis is pointed towards a position angle (PA)
    of 35\arcdeg. The NIC3 images have been scaled to the
    0.0996\arcsec/pixel scale of the WFPC2 camera.}
\end{figure*}

\subsubsection{ACS Images of a Southsoutheast Field Previously Observed with NICMOS in \htwo}
A field to the southsoutheast of the central star and falling into the outer-ring portion of the Helix Nebula
was imaged with the ACS camera. Eight exposures of nominally 1200 s each were made in both the F658N filter
(which passes the \Halpha\ and [N~II] 6583 \AA\ lines equally well) and the F502N filter (dominated by the [O~III] 5007 \AA\ line). The field of view shown in Figure 1  overlaps only slightly with the ACS mosaic built up
during the GO 9700 survey with the same filter pairs (OMM04). The signal to noise ratio was much higher than in the
GO 9700 survey since that study used total exposures of about 850 s in both of filters.

The images in the four offset pointings were combined using tasks within the STSDAS package. 
No attempt at absolute calibration was made because of the F658N filter requiring additional observations with the
F660N filter, which is dominated by [N~II] (O'Dell 2004). The resulting image for F658N is shown in Figure 3. Originally of 0.05\arcsec\ pixels, it has been averaged into 2x2 samples in order to increase the 
signal to noise ratio. Since the finest features are larger than the size of the
resultant pixels (0.1\arcsec) no loss of detail was incurred. The F502N image is much lower signal and is not presented
here, although portions of it are reproduced and discussed in \S\ 3.4. 

\begin{figure*}\centering
  \includegraphics{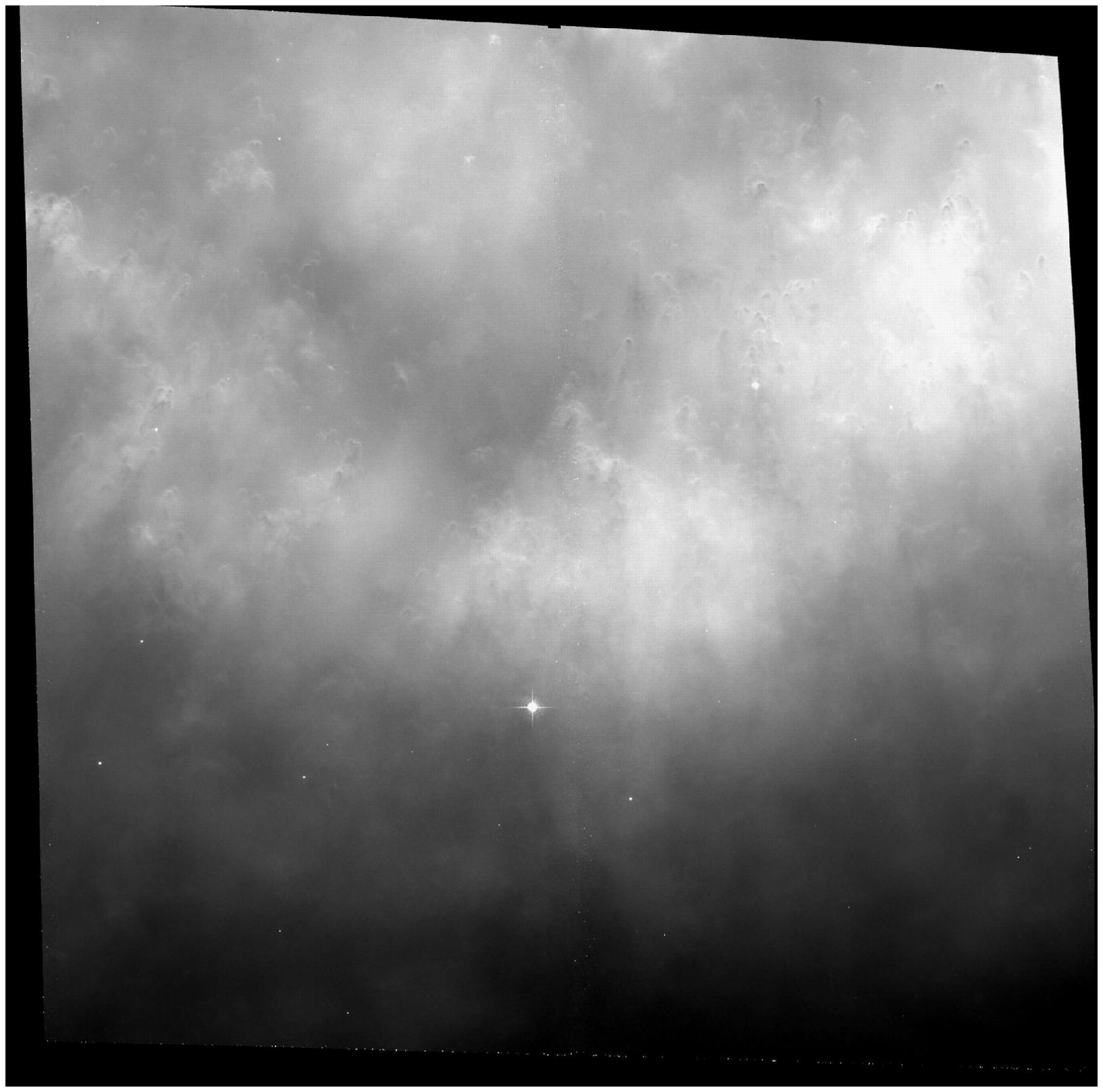}
  \caption{This high contrast 215\arcsec x 213\arcsec\ image of the
    ACS region shown in Figure 1 has the vertical axis pointed towards
    PA=350\arcdeg. It demonstrates that bright cusp knots are found in
    the outer-ring of the nebula, in addition to the inner-ring knots
    that dominate the field covered by the GO 9700 mosaic.}
\end{figure*}

\subsubsection{WFPC2 Images in the He~II filter of a Southwest Field}
Sixteen exposures of 1100 s each were made with the WFPC2 F469N filter that isolates the He~II recombination
line at 4686 \AA. The images from the four pointings were combined using the STSDAS package of {\it dither} tasks
and the results are shown in Figure 4. For comparison, a matching section of the ACS mosaic that was derived as part 
of program GO 9700 (OMM04) is also shown.

\begin{figure*}
  \includegraphics{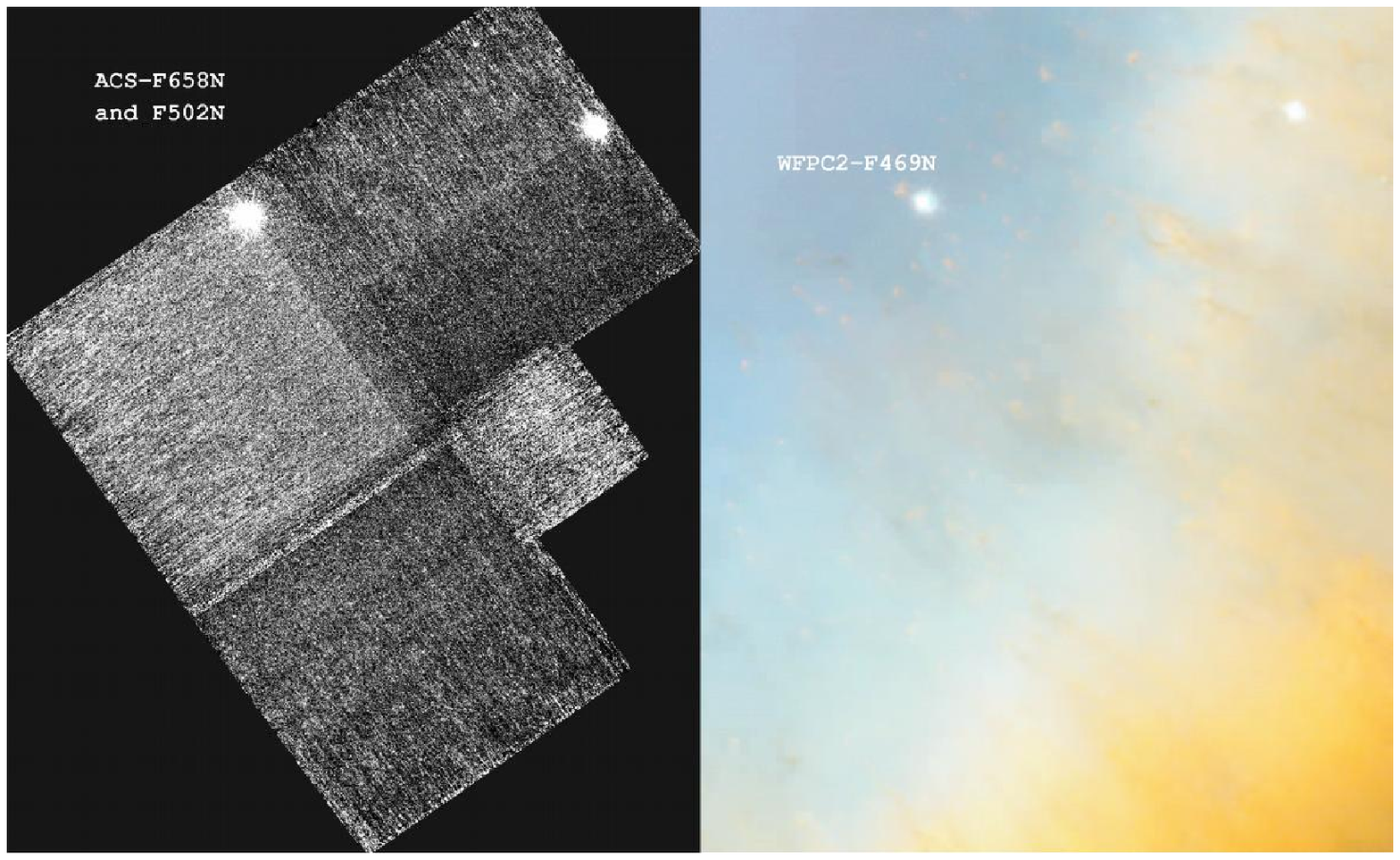}
  \caption{The left panel shows a 167\arcsec x 204\arcsec\ WFPC2 F469N
    image with north as the vertical axis and is the result of
    combining 16 dithered images.  For comparison,the right hand panel
    shows the same section of the nebula from a mosaic of ACS images
    (OMM04) with \Halpha +[N~II] as red, an average of \Halpha +[N~II]
    and [O~III] as green, and [O~III] as blue.  No indication of
    features showing HeII 4686\AA\ emission is seen.}
\end{figure*}

\subsection{Recently Published Observations}
There are three recent papers that contain observations pertinent to our discussion of the nature of the 
knots in the Helix Nebula. These include two space infrared observations and one groundbased study.

\subsubsection{Spitzer Space Telescope Infrared Images and Spectra}
In a recent paper H06 present the results of an extensive observational study of the Helix Nebula. They have imaged 
the object out to the northeast-arc (OMM04) with the IRAC camera (Fazio et~al. 2004) in four broad filters centered on 3.6 $\mu$m, 
4.5 $\mu$m, 5.8 $\mu$m, and 8.0 $\mu$m.  These filters are dominated by emission from rovibrational lines within
the $v=0$ ground electronic state, although an atomic continnum must be present in addition
to a few collisionally excited forbidden and recombination emission lines. The resolution of these images is about 
2\arcsec, so that one cannot resolve structure within the cusps, but one can see structure along radial lines passing
through the bright cusps and their much longer tails. Spectra were obtained with the IRS (Houck et~al. 2004) at 
three positions, two falling in the outer-ring at locations north and southwest of the central star and the third
location (used for background subtraction) falling directly on the fainter northeast-arc feature. They present
calibrated fluxes for rovibrational lines of \htwo\ from 0-0 S(7) at 5.51 $\mu$m out to 0-0 S(1) at 17.0 $\mu$m.
They also present a groundbased image in a filter centered on the \twomic\ \htwo\ line which is of comparable
spatial resolution but wider field of view that the Speck et~al. (2002) study. This image is not
flux calibrated but appears to go fainter than the Speck et~al. (2002) image.

\subsubsection{HST GO 9700 images in \htwo}
In the program GO 9700 study that produced a continuous mosaic of ACS images MX05 also obtained parallel NIC3 images 
in the F212N filter at six science positions and one sky position. Each position was actually a double exposure
of two pointings, which allowed a slight overlap of the NIC3 fields. The total exposure at each position was about 
750 s and double that in the regions of overlap.  The method of calibration was different as only F212N images
were obtained (the few F175W images were not useful for calibration of the F212N images). It was assumed that the 
sky images included all the background signal that need to be subtracted, which means that it did not subtract nebular
continuum.  The NIC3 images are under-sampled (the pixel size of 0.2\arcsec\ is about the same as the telescope's resolution at this wavelength) as in our new observations of the field around 
378-801. When comparing the data, one should note that the maximum effective exposure time is about 1500 s
for the GO 9700 F212N images, while in our study it is 10,240 s. As discussed in \S\ 3.4, our ACS field 
overlaps with much of the MX05 position 2 field, allowing a more meaningful comparison of optical and infrared images
of this region than was possible in the MX05 study.

\section{Analysis of the Observations}
These new images of the Helix Nebula with three different HST cameras provide new information on a number of subjects related to the nature and formation of the knots. The NICMOS \htwo\ image allows a discussion of the cusp and tail structure, the He~II 
image places constraints on the location of the knots to the southwest, and the ACS images allow a more complete comparison of
early \htwo{} images to the south and southeast with comparable resolution optical emission line images.

\subsection{Stratification in the Bright Cusp of 378-801}

The new NIC3 \htwo\ \twomic\ line observations were scaled to the same pixel size as the WFPC2 (0.0996\arcsec
/pixel) using bilinear interpolation and carefully aligned. Although this region has been imaged numerous times in this line, this is clearly the 
best image in terms of both its resolution and signal. The three primary optical line images and the new \htwo\ image
are shown in Figure 2. In order to make a quantitative analysis, the images were rotated and a sample three pixels wide along
the axis of the cusp-core-tail was made. The results are shown in Figure 5, where the peak value of the \htwo\ \twomic\ and [NII] emission is
normalized to unity as are the outer portions of the [OIII] profiles.

\begin{figure}
  \includegraphics{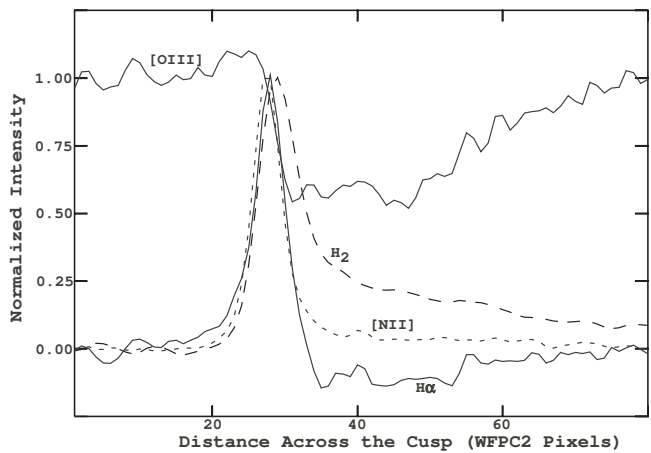}
  \caption{The normalized intensity of a sample along the axis of knot
    378-801 is shown for three optical lines and the \htwo\ \twomic\
    infrared line. The \twomic\ image has lower resolution (about two
    pixels) and a comparison of the appearance in the different lines
    is discussed in the text.  }
\end{figure}

The optical line results are similar to those found by O'Dell \&\
Burkert (1997), Burkert \&\ O'Dell (1998), O'Dell, Henney, \&\ Burkert
(2000) and OHF05 in that the ionization occurring furthest from the
knots ionization front ([O~III]) is weak and extended, this intrinsic
low brightness allows one to see the core of the knot in silhouette
against the background nebular emission in this line. Although the
extinction peaks in the core of the knot, it extends to about
5\arcsec\ from the bright cusp.  The [N~II] emission is strong and
displaced (0.05\arcsec) away from the ionization front with respect to
the \Halpha\ emission.  One sees extinction in \Halpha\ from the core
out to almost the same distance as in [O~III]. The lack of apparent
extinction in [N~II] must be due to there being relatively more
emission in that line in the sheath of ionized gas surrounding the
shadow of the knot. The full width at half maximum (FWHM) of the
\Halpha\ and the [N~II] images are 0.48\arcsec\ and since the FWHM of
the stars in the field of view is 0.25\arcsec, quadratic subtraction
of this instrumental component leaves an intrinsic line width in those
emission lines of 0.41\arcsec. The FWHM of the \htwo\ line is
0.61\arcsec\ and the nearby star's is 0.42\arcsec, leaving an
intrinsic FWHM for \htwo\ of 0.44\arcsec\ with a peak displaced
0.11\arcsec\ towards the core of the knot from the \Halpha\ peak. The
FWHM corresponds to a length of $1.40 \times 10^{15}$ cm and the displacement
to $3.5 \times 10^{14}$ cm.  The ionized line characteristics are similiar to
those found in the previous studies, but we have added here the
important characteristic of the small but certain displacement of the broader
\htwo. The earlier \htwo\
studies lacked the resolution to determine this characteristic, or in
the case of MX05, lacked the high resolution optical lines necessary
for the comparison.

The calibrated peak surface brightness in the cusp of 378-801 is $1.8 \times 10^{-4}$ \sbunits\ in the \htwo\ \twomic\ line.
This agrees well with the peak value of $1.0 \times 10^{-4}$ \sbunits\ found by Huggins et~al. (2002), where they used the calibration of
Speck et~al. (2002) and utilized a spatial resolution that would not have recognized the narrowness of the peak.

\subsection{Stratification in the Tail of 378-801}

The well defined tail in 378-801 is primarily formed by a shadow in the ionizing Lyman Continuum (LyC) radiation cast by the optically
thick core, with illumination occurring by diffuse (recombination) LyC photons and direct radiation grazing the 
edge of the core.  The first order theory describing this situation was presented by Cant\'o et~al. (1998) and applied to
the tails of the Helix Nebula and the shadows behind the Orion Nebula proplyds soon after (O'Dell 2000).  
OHF05 discussed the structure in tail of 378-801 within the light of this theory and its next order refinements (Wood
et~al. 2004), but were unable to explain the details of what was being seen.
\htwo\ in the tails was first detected by Walsh \&\ Ageorges (2003) and we are able to establish with our new observations
where this emission arises.

We present in Figure 6 results from traces across the tail of 378-801 extending from 3.1\arcsec\  to 6.0\arcsec\ behind
the peak of the cusp in \Halpha. This region does not extend as far as the partially obscured knot lying on the
east side of the tail with its cusp 8.3\arcsec\ beyond 378-801's \Halpha\ cusp. One sees 
that there is a well defined signature of a limb brightened sheath in both \htwo\ \twomic\ and \Halpha, with the peaks of the 
\htwo\ emission occuring inside the \Halpha, as expected if the \Halpha\ is associated with a local 
ionization front. Since the gas ionized by diffuse radiation should be much cooler than the directly illuminated
nebular gas, the \Halpha\ emissivity would be high and the ionized sheath is well defined. These observations establish that 
conditions in the tail do allow an ionization front to form, while Cant\'o et~al. (1998) and O'Dell (2000) had 
expected that the shadowed region may be fully ionized.  It is not surprising that no [O~III] emission is seen,
rather, that the dust in the tail, concentrated to the middle of the tail (as noted by OHF05), causes extinction
of the background nebular light.  The apparent quandary, noted in OHF05, is that the [N~II] emission appears to
come from inside the ionized sheath of the tail. We already noted that the \Halpha\ and \htwo\ structure indicates
that an ionization boundary occurs at the edge of the radiation shadow, so there is an apparent contradiction in
finding ionized nitrogen emission originating inside an ionization front. 

\begin{figure}
  \includegraphics{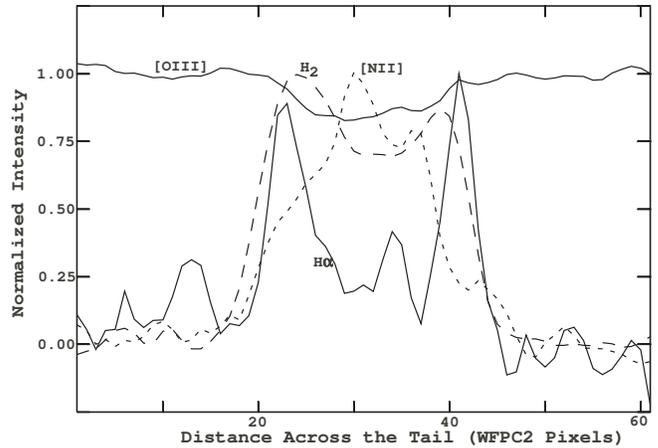}
  \caption{A cross-section of a tail region of knot 378-901 is shown,
    with the radiation in each line normalized to unity.  The details
    of the sample and interpretation are presented in the text.}
\end{figure}

This contradiction is removed by comparison of the [N~II] emission and the optical depth, as determined by
the [O~III] image.  Figure 7 shows a comparison of the dust optical depth and the [N~II]
intensity. The similarity of the distribution of the optical depth and the [N~II] brightness argues that the
[N~II] is actually nebular or cusp light scattered by the only marginally optically thick (peak value $\tau$=0.2 column
of dust). This feature is easy to see because the expected low electron temperature of the sheath's ionization front
would suppress the collisionally excited [N~II]. If our intepretation of [N~II] is correct, then there should be
a similar component of scattered \Halpha\ radiation. This may be what somewhat fills-in the region between the
two limb-brightened components (in addition to the low level of surface brightness expected when examining a
thin shell).  In the cusp [N~II] is stronger than \Halpha\ emission whereas in this part of the nebula the opposite is true. This point argues
that much of the scattered [N~II] emission arises from the nearby bright cusp, rather than the surrounding nebula.

\begin{figure}
  \includegraphics{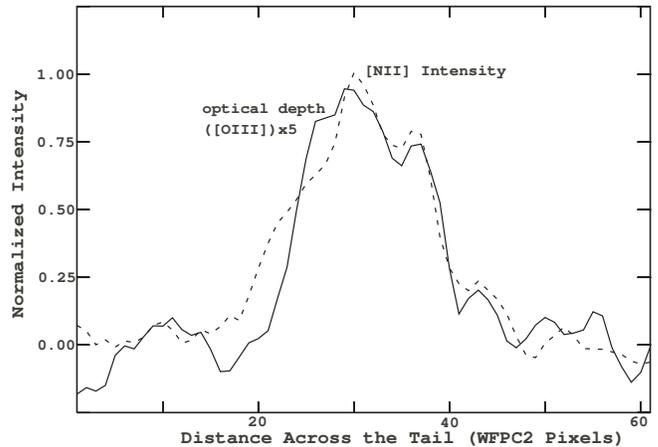}
  \caption{The normalized surface brightness of the tail sample of
    Figure 6 is shown together with the optical depth (multiplied by
    five) determined from the [O~III] profile. The similar
    distribution of each argues that the [N~II] emission is caused by
    scattering of nebular [N~II] emission, as discussed in the text.}
\end{figure}

The distribution along the tail core is discussed in detail in OHF05 (their \S\ 4.1.2) and only a few comments
need be added.  The question is complex because the knots seem to originate near the nebular ionization front,
then are shaped by the radiation field as the ionization front expands beyond them (O'Dell et~al. 2002). This means
that material in the shadowed region will have never seen direct ionizing radiation and could represent pre-knot
material from the planetary nebula's  photon dominated region (PDR). The other source of tail material could be neutral gas accelerated
outwards by the rocket effect (e.g. Mellema et~al. 1998). Unfortunately, the high velocity resolution 
study of CO by Huggins et~al. (2002) does not really illuminate the question. Their angular resolution was a Gaussian
beam of $7.9\arcsec \times 3.8\arcsec$, with the long axis aligned almost along the axis of the tail of 378-801.  
Since there was a strong CO component coming from the partially shadowed knot lying 8.3\arcsec\ beyond 378-801's
\Halpha\ bright cusp, this means that there is a not a clear resolution of the CO emission from the core of
378-801 and the partially shadowed knot.  As a result, one cannot hope to interpret the small differences in the 
position of the peak emission at different velocities as core-tail differences. This could be done with higher spatial resolution CO observations.

\subsection{An Unsuccessful Search for He~II emission in the Southwest Knots}
Figure 4 shows our deep HeII images alongside the same field covered at comparable resolution with the ACS in \hanii\ and 
[O~III]. A detailed comparison of the two images indicates that there is no case of a HeII feature corresponding
to a \hanii\ or [O~III] feature, nor any HeII only features.  The part of the WFPC2 field closest to the central
star is 144\arcsec\ distance.  The profile of the HeII core of the central disk, to which the knots in this part
of the nebula belong (OMM04) derived from a wide field of view HeII image (O'Dell 1998) shows that the core is down
almost to 50\%\ of its peak emission at this distance. If any knots actually occur within the nebula's HeII core, we would expect that in the simplest knot models, we would see a HeII cusp outside the [O~III] zone of each knot and this is not the case.  However, the basic
model is not that simple. 

The detailed models of OHB00 and OHF05 show that the normal progression of ionization states in the cusp are 
preserved, that is, closest to the ionization front there is an He$\rm ^{o}$+H$\rm^{+}$ zone, outside of which there 
is a He$\rm ^{+}$+H$\rm^{+}$ zone, and outside of that a He$\rm ^{++}$+H$\rm^{+}$. In a nearly constant electron
temperature nebula the innermost zone is best traced by the [N~II] emission, the next zone by the [O~III] emission, 
and the outermost zone by the HeII emission. 
Things are not so simple in the case of the knots. As the gas flows through the cusp it is heated only slowly, so that
the collisionally excited [N~II] emission peaks further out, where the gas is hotter, more than making up for the 
lower fraction of N$^{+}$ ions. By the time that the second zone is reached the density has dropped significantly and
the [O~III] emission is broad and weak. One would expect to find a HeII zone associated with a knot only If the knot lies within the nebula's HeII core. 
This HeII zone would be quite weak because the density of the knot's gas would have been greatly decreased this far out.  Moreover, the gas has
probably nearly reached the temperature of the nebula and these higher temperatures suppress the emission of this
recombination line. This means that it will be hard to actually detect by their HeII emission any knots within the HeII core of the nebula.
Probably the strongest evidence that no knots exist in the HeII core lies in the fact that we don't see any objects in extinction in any of the observed
emission lines.

\subsection{Comparison of the ACS Images in the Southsoutheast with earlier NICMOS \htwo\ Images}

As noted in \S\ 2.1.2, our ACS field overlapped with one of the double-pointings made with NIC3 in F212N as part of
program GO 9700 (MX05). MX05 compared their F212N images with the corresponding five fields in OMM04, where the
resolution was about 1\arcsec and groundbased images were used because the HST ACS mosaic did not extend out this far.  This factor of five difference in resolution made it difficult to draw firm 
conclusions about differences and similarities of appearance. They did, however, conclude that even their short
(750 s) overlapping double exposures were sufficient to establish that the knot cusps were more visible in \htwo. In \S\ 2.1.1 we
showed that in the vicinity of 378-801 that the knots are equally visible in both the ionization cusps and the
\htwo\ cusps. 

In Figure 8 we see a comparison of the GO 9700 F212N (\htwo) images with our new ACS images.
There is an excellent correlation of appearance, although the contrast of the \htwo\ emission above the 
essentially zero nebular background is higher than in F658N (\hanii) and as usual the knots are only easily seen in
the F502N ([O~III])  when the knot is in the foreground and can be seen in extinction against the background
nebular emission. After considering the flexibility of display of the high signal to noise \hanii\ images, it is difficult to support the conclusion that \htwo\ \twomic\ images are a better way of
searching for knot cusps, except for any regions where there is high obscuration. 

\begin{figure*}
  \includegraphics{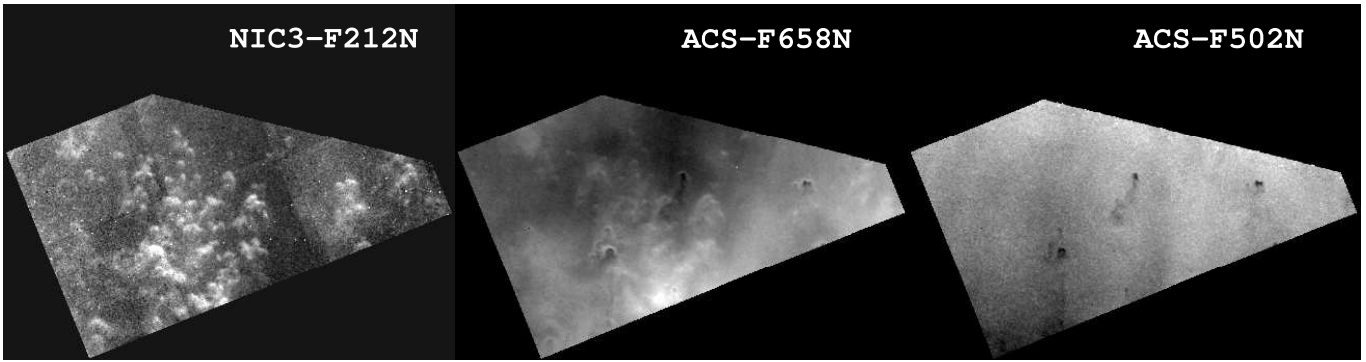}
  \caption{This figure shows the appearance in the 88\arcsec\ wide
    field of overlapping ACS (GO 10628) and NIC3 (GO 9700) images,
    with north up. The \htwo\ cusps of the knots are much more visible
    than the combined \hanii\ ionized cusps. However, close
    examination shows an ionized gas feature corresponding to each
    \htwo\ feature.  This is not true for the [O~III] F502N features,
    which require that the knot be located in the foreground. The
    faint circular features are artifacts of the imaging system.}
\end{figure*}

\subsection{Variation in the \htwo\ Cusps with Distance from the Central Star}

Through the new \htwo\ \twomic\ NIC3 images in the current program (GO 10628) and the earlier MX05 study with
shorter exposures, there is a larger and hopefully representative sample of resolved \htwo\ cusps over a wide 
range of stellar distances ($\phi$).  To look for systematic differences, we have identified isolated cusps 
in each of the fields available. We selected the three closest knots in the GO 10628 field, three in both of
MX05's positions one and two, and two in MX05's positions three and four, no isolated cusps being available in MX05's
position five. In each case we measured the surface brightness at the peak of the \htwo\ cusp, the approximate
chord across the knot's center, the width of the \htwo\ cusp, and determined $\phi$.  
The GO 10628 and MX05 positions 1-4 had average $\phi$ values of 139\arcsec, 290\arcsec, 278\arcsec,
375\arcsec, and 464\arcsec\ respectively.  The average surface brightnesses (in \sbunits) of the cusp peaks were 
$1.37 \times 10\rm ^{-4}$, $7.5 \times 10\rm ^{-5}$, $8.5 \times 10\rm ^{-5}$, $4.9 \times 10\rm ^{-5}$, and $4.5 \times 10\rm ^{-5}$. The average cusp widths
were 0.6\arcsec, 0.3\arcsec, 0.5\arcsec, 0.8\arcsec, and 1.1\arcsec. The average chord values were 
1.5\arcsec, 2.5\arcsec, 2.2\arcsec, 2.4\arcsec, and 2.7\arcsec. 

The most pronounced change in knot characteristic is the cusp peak surface brightness, dropping about
linearly with $\phi$, with the cusp width growing slowly and the chord width more steadily with $\phi$. The
physical interpretation of these patterns awaits a better understanding of individual knots. It should be pointed
out that the relative physical distances from the stars is likely to be increasing more rapidly than the relative values of $\phi$. This is because in the 3-D model of the Helix Nebula in OMM04, the main disk is inclined
at an angle about 23\arcdeg\ out of the plane of the sky, with the northwest side closer to the observer, while the
outer-disk is inclined about 53\arcdeg\ out of the plane of the sky, with the southsoutheast side closer to the observer.
Objects associated with the inner-disk would have a distance multiplication factor of 1.09 and those in the 
outer-disk a multiplication factor of 1.66. The objects in the GO 10628 NIC3 field are almost certainly associated
with the inner-disk. MX05's positions 1 and 2 could belong with either system (accurate radial velocities would
determine this) and their positions 3 and 4 are almost certainly associated with the outer-disk.


\subsection{Comparison of the Structure of the Knots in 2.12 $\mu$m and in \Halpha}

Figure 2 demonstrates the remarkable similarity of appearance of the knots in \Halpha\ and in our F212N (\twomic) images
and we discussed the quantitative properties of 378-801 in \S\ 3.1 and \S\ 3.2. We have investigated the similarities of the knots by
selecting the nine objects (including 378-801) within the NIC3 field of view that are sufficiently isolated to allow a good background subtraction.
The peak surface brightness in each cusp was derived in both \twomic\ and \Halpha\ for a sample 3 WFPC2 pixels wide (0.3\arcsec).
The peak surface brightness in the tail of each object was determined in a sample 11 WFPC2 pixels wide (1.1\arcsec) across the
tail, the closer end of the sample being 20 WFPC2 pixels (2.0\arcsec) displaced from the tip of the bright cusp. We used the
\twomic\ calibration described in \S\ 2.1.1 and the \Halpha\ calibration of O'Dell \&\ Doi (1999), expressing the surface brightnesses in
units of photons cm$^{-2}$ s$^{-1}$ sr$^{-1}$. The average cusp surface brightness ratio in \twomic /\Halpha\ was 5.5$\pm$1.0.
The average surface brightness ratio for the tail as compared with the cusp was 0.23$\pm$0.08 in \twomic\ and 0.17$\pm$0.05 
in \Halpha.  This means that in the cusp the \htwo\ \twomic\ line alone is putting out more than five times as many photons as 
in \Halpha\ and that the contrast between the tail and cusp may be slightly higher (the ratio is lower) in \Halpha\ than in \twomic.
The remarkable similarity of the knots in a recombination line that follows photoionization of atomic hydrogen and emission in
a molecular line heretofore assumed to be the result of radiative pumping between electronic states is discussed in \S\ 4.5.

\section{Discussion}
Our understanding of the physics of the knots has evolved with a better understanding of a model that satisfactorily
explains the knots.  The initial descrepancy between the cusp surface brightness and the 
simplest photoionization model (O'Dell \&\ Handron 1996) was resolved by L\'opez-Mart\'{\i}n et~al. (2001) when it was shown that 
the advection-dominated nature of the flow through the knot ionization fronts leads to a total rate of recombinations in the ionized gas that is significantly below what is predicted from naive models of ionization equilibrium. The peculiar photoionization structure of \Halpha\ and [N~II] emission can also be understood in similar terms---the heating timescale of the ionized gas is comparable with the dynamic timescales for flow away from the knot surface, leading to resolvable temperature gradients, which strongly affect the relative distribution of recombination line and collisional line emission. The most refined model is that of OHF05, which included both the effects of the radiation field and also the hydrodynamic expansion of the knot's ionization front. It is probably accurate to say that the photoionized portions of the knots are now adequately understood, or at least that the models are broadly consistent with the best observations.

The structure in the tails is only beginning to be understood. To the first order, the tails are the effects
of radiation shadows in the dominant ionizing species, the LyC photons (Cant\'o et~al. 1998, O'Dell 2000). With this paper (\S\ 3.2)
we have now determined that the well observed tails are ionization bounded, with \htwo\ sheaths inside the
zone of ionized gas that occurs at the edge of the LyC shadow. The inner part of the tail is dense enough in
dust to scatter surrounding nebular light, although the origin of this material as arising from the original
process that forms the knots or as material that is moving back from the knot remains uncertain.

The greatest quandary surrounds the explanation of the \htwo\ zone that is observed immediately inside the 
ionized cusps of the knots.  The approximate location of this zone of observed \htwo\ is qualitatively where 
one would expect it. For reasons given below, it is almost certainly not excited by shocks. Other models (e.g. Natta \&\ Hollenbach 1998) argue that the heating
is by absorption of soft X-rays and others that the 
excitation mechanism is probably fluorescence, where non-ionizing photons from the stellar continuum
excite molecules to the B~$\rm ^{1}\Sigma ^{+}_{u}$  and C~$\rm ^{1}\Pi _{u}$ electronic states, which then decay, producing the populations of the ground electronic
state that give rise to the observed infrared lines. Within the core of the knot the density is sufficiently high
and the temperature sufficiently low that multiple heavier molecules are formed and the observed CO is simply the most
easily observed abundant tracer of these heavy molecules.

An alternative method of exciting the \htwo\ molecules is by shocks. At first this idea seems attractive because
planetary nebulae as a class are known to possess high velocity stellar winds and large scale mass flows
with sufficient energy to excite the low lying energy states of \htwo\ that give rise to the observed infrared
lines. Cox98 first pointed out that the lack of a stellar wind (Cerruti-Sola \&\ Perinotto 1985) rules out excitation by wind-driven shocks.
A more complete assessment of shocks as the exciting source was given in OHF05 (their \S\ 4.3.2), where it is
shown that although \htwo\ is heated sufficiently immediately behind a transient shock  this shock would quickly move  
through the knot and up the tail. The well defined location of the \htwo\ emission zones immediately behind the
ionized cusps and the ionized sheath of the tail strongly argues that we are dealing with a quasi-stationary
process, rather than something quite dynamic, like shocks. 

H06 base their interpretation of \htwo\ emission on the assumption of shock excitation.  Their assumption
is based on the weakness of the \htwo\ 1-1 S(7) line, stating that the radiative models they use predict strong
emission in that line, which they do not observe in their spectra. The shortcomings of that criterion for
determining that the excitation comes from shocks, rather than radiative processes is discussed below in
\S\ 4.1.  As we show in \S\ 4.2, the H06 spectra also argue for a high excitation temperature, as found by
Cox98. The H06 shock interpretation of the relative population distribution of the \htwo\ energy states used six 
free parameters as it required three different shock velocities, each with a different relative intensity. This means
that one can't use the population distribution to confirm that method of excitation.

A key element of understanding the \htwo\ emission is the excitation temperature of the gas. Cox98 used spectra
of the \htwo\ 0-0 S(2) to S(7) lines to derive the population of their upper states and found that their two
sampled regions matched an excitation temperature of 900$\pm$50 K. We show below (\S\ 4.2) that the new spectra
of H06 of the \htwo\ 0-0 S(1) to S(7) lines in two additional regions agree with the results of Cox98 and support the idea that the
\htwo\ emission comes from gas that is much hotter than the 50 K conditions expected (OHF05) in the core of the knots.

OHF05 demonstrated that even their most detailed PDR models could not explain the high surface brightness in \htwo\ 
of the 
knot cusps, an argument first made by Cox98 from more general considerations.  The argument reduces to the fact
that the observed surface brightness in \htwo\ \twomic\ radiation is too high to be explained by the column
density of 900 K \htwo\ that is predicted. 
OHF05 did not have high resolution \htwo\ \twomic\ images of their sample knot (378-801) and a comparison using
the results of the new observations reported here are given in \S\ 4.5.

Several papers, including the recent MX05 study, have reported that the surface brightnesses are compatible with 
earlier the theoretical models of Natta \&\ Hollenbach (1998) in spite of the fact that those authors point
out that the knots do not adhere to their general model and would have a higher surface brightness. A more 
complete critique of earlier claims of agreement of theory and observations is given in OHF05 (\S\ 4.3.3).

In this section we present the total flux from the central star and nebula in \S\ 4.1,
establish that the knots commonly have high excitation temperatures (\S\ 4.2),
show that the absence of strong 1-1 S(7) emission is not a strong argument for shock excitation of the \htwo\ (\S\ 4.3),
compare the recent data on \htwo\ emission with the best models (\S\ 4.4), determine that there
is no evidence for radial features extending into the middle of the nebula (\S\ 4.5), 
and critique a recent paper that argues for the tails being formed primarily by hydrodynamic processes in \S\ 4.6.

\subsection{The Total Flux from the Helix Nebula and its Central Star in Various Energies}

The emission from the nebula is in at least quasi-equilibrium with radiation from
the central star. This means that the relative fluxes in various nebular and cusp emission lines and in the stellar continuum
impose important constraints that must be observed by the correct model for the cusp \htwo\ emission.

\subsubsection{The Central Star}

The stellar continuum has been well defined down to 1200 \AA\ by Bohlin et~al. (1982), who conclude that the star
has a luminosity of $120~L_{\odot}$ (corrected to the trigonometric parallax distance) and an effective temperature of 123,000~K.
In the long wavelength end of the continuum, the flux per wavelength interval is very close to $\lambda^{-4}$, as expected when
one looks at much lower energies than where the peak emission occurs. This total luminosity corresponds to a flux at the
Earth of $8.8 \times 10^{-8}$ \funits. 

Natta \&\ Hollenbach (1998) argue that the \htwo\ is heated by X-rays of greater than 100 eV because only these high
energy photons would penetrate the ionization boundary. There are two emitters in the high energy end of the spectrum,
the central star and a high temperature component of about (10$^{7}$ K) (Leahy et~al. 1994, Leahy et~al. 1996, Guerrero et~al. 2001). The central star emission in the 0.1-2.0 KeV range is
$4 \times 10^{-11}$ \funits\ and the emission from the 10$^{7}$ K component is $9 \times 10^{-14}$ (Leahy et~al. 1994). 

The wavelength range for the 
fluorescent pumping mechanism is from about  912 \AA\  to 1100 \AA\ as determined by the minimum energy for exciting the
Lyman bands and the cutoff imposed by the LyC absorption of hydrogen. Extrapolating the continuum from the slightly longer
wavelengths that have been observed gives a total flux in this interval of $6.9 \times 10^{-9}$ \funits. If the heating is due to 
photons above the ionization threshold for neutral hydrogen, then the calculated flux for a 123,000 K blackbody of 120 L$_{\odot}$
in the interval from 13.6 eV through 100 eV is $8.1 \times 10^{-8}$ \funits, representing the largest amount of power coming from the central star. The observed and predicted properties of the star's flux are summarized in Table 1.
\begin{table}[t]
  \caption{Nominal fluxes at earth from the Helix central star$^a$}
  \smallskip
  \label{tab:fluxes}
  \begin{tabular}{lr@{\hspace*{4em}}lll}
    \toprule
    & &  \multicolumn{3}{c}{Flux (erg cm$^{-2}$ s$^{-1}$)} \\
    \multicolumn{2}{l}{Wavelength range} & Model$^b$ & Black Body & Observed \\ \midrule
    X-ray & $< 124$~\AA  & $4.2\times 10^{-11}$ & $1.1\times 10^{-9}$  & $4\times 10^{-11}$ \\
    EUV & 124--912~\AA   & $8.3\times 10^{-8}$  & $8.1\times 10^{-8}$  & \nodata  \\
    FUV &  912--1100~\AA & $1.9\times 10^{-9}$  & $2.3\times 10^{-9}$  & $6.9\times 10^{-9}$  \\
    \bottomrule
  \end{tabular}\\[\medskipamount]
  $^a$Assuming no intervening absorption and $L = 120~L\sun$, $T_\mathrm{eff}=1.23\times 10^5$~K, $D=213$~pc.\\
  $^b$Model fluxes are from the $\log g = 7$, solar abundance model of Rauch (2003).
\end{table}

\subsubsection{The Nebula's Emission Line Flux}

The flux from the entire nebula was determined by O'Dell (1998) to be F(\Hbeta)=$3.37 \times 10^{-10}$ \funits\ and
F([O~III] 5007 \AA)=$1.94 \times 10^{-9}$ \funits\ from narrow band filter images. The proximity of the nebula and
its line ratios indicate that the interstellar extinction is low and little correction is necessary.

We have determined the total flux in the \twomic\ line using the calibrated \twomic\ image of Speck et~al. (2002). Stars were edited-out by 
hand, with the local values substituted, and the background was assumed to be found in the west of their field, a region of much lower
nebular surface brightness in optical lines (OMM04). This process gave F(\twomic)=$6.2 \times 10^{-10}$ \funits. Using the \Halpha\ to 
\Hbeta\ flux ratio (2.79) of O'Dell (1998), indicates that the number ratio of \twomic\ to \Halpha\ for the nebula as a whole
is 2.13, which is less than half the ratio of 5.5 for the cusps in the NIC3 field of view. This is consistent with the 
fact that almost all of the \htwo\ emission arises from the cusps, rather than the nebula.

Cox98 have estimated the total flux in the \htwo\ lines falling into their LW2 image to be $2.5 \times 10^{-9}$ \funits.
This image contains the lines in the 0-0 S(4) through S(7) series. The spectra of Cox98 and H06 indicate that the 
strongest line, which is 0-0 S(5), is 52\%\  of the total flux of these lines, so that the 0-0 S(5) line has a total flux
from the nebula of $1.3 \times 10^{-9}$ \funits. If one assumes the line ratios of the H06 study, then the total emission
from the nebula in the 0-0 S(1)-S(7) lines of \htwo\ becomes $3.8 \times 10^{-9}$ \funits. If one includes the \twomic\ emission line,
this means that the total observed \htwo\ flux is $4.4 \times 10^{-9}$ \funits\ and the total \htwo\ emission must be more because there are
numerous transitions that have not been observed. 

A comparison of the optical recombination plus collisionally excited lines with the \htwo\ emission indicates that comparable amounts of 
radiation from the nebula is coming out as \htwo\ emission. Since no perceptible \htwo\ emission comes from the nebula, this means 
that the process producing the \htwo\ emission in the knots is working very efficiently.

Any mechanism seeking to explain the \htwo\ emission must account for
$\geq$ $4.4 \times 10^{-9}$ \funits. This value is much larger than the soft X-ray stellar flux of $4 \times 10^{-11}$ \funits . The \htwo\ flux is comparable to the 912 -1100 \AA\  total stellar flux 
of $6.9 \times 10^{-9}$ \funits. However, since the fluorescence mechanism operates by absorption of relatively narrow samples within this wavelength range,
it appears that this mechanism too cannot provide enough energy to power the \htwo\ emission.  There is certainly enough power available
in the stellar continuum, if the mechanism depends upon absorption of a broad wavelength range, rather than narrow emission lines. This means that $\geq$64\%\ of the
photons with energies less than 13.6 eV would need to be absorbed or about $\ge$5\%\ of the 13.6-100 eV radiation. These fractions would have to be larger if the knots are concentrated to a small fraction
of the view of the central star. This is certainly the case as the knots are exclusively found in the lower
ionization portions of the main disk of the nebula and the inner boundary of the outer-disk (OMM04).
This consideration finally rules out any possibility of the non-ionizing photons as a source of the
power.

\subsection{The population distribution of the levels producing the observed \htwo\ lines}

The nature of the population distribution in the upper states that produce the observed infrared \htwo\ emission lines can be
a powerful diagnostic in understanding the physical procedures operating in the knots. Therefore, we have determined this distribution
using the two available sets of data.

Cox98 demonstrated that the spectrum in both of their samples closely matched a single excitation temperature of  900$\pm$100 K, using this
result to argue that the \htwo\ regions emitting the 0-0 S(2) to 0-0 S(7) lines had total densities $\geq 10^{5}$ \cmq. The Cox98 samples
are in the region called the outer-ring by OMM04. H06 observed 
over a slightly larger wavelength range, reporting the detection of the 0-0 S(1) line at one of their two positions, but did not present
a derived population distribution, probably because they assumed that the population was determined by shocks, as discussed in the 
next section. H06 does not identify where the samples were obtained except that they were in the ``main ring of the nebula." An examination of 
the Spitzer Space Telescope data base show that both H06 samples also fall in the region of the outer-ring at distances from the central star
that are very similar to those of Cox98.

We have determined the population distribution for each of the four positions for which these studies provide data. H06 present tabulated line
fluxes, including the same lines as the Cox98 study with the exception of reporting a flux for the 0-0 S(1) line at one of their positions. A plot
of their spectra does not include this wavelength region. No errors are reported for their fluxes and it is impossible to judge the accuracy of the single
0-0 S(1) flux without a spectrum, so that we will not use that line in our analysis. Cox98 did not give tabulated values of their fluxes, however, they
too presented plots of the spectra for both regions. These plots were used to measure the line fluxes relative to the strongest line 0-0 S(5).
Because of the importance ascribed  by H06 to the presence or absence of the 1-1 S(7) line, we carefully looked at its expected wavelength and made
a very marginal detection in Cox98's more southern position. Because the uncertainty of that flux measurement is so great, we'll not use it in this analysis of the 
population distribution.

We have added one additional point in this study by including the \twomic\ line, which is the 1-0 S(1) line. Since there is not a 
matching sample of the nebula the \twomic\ and 0-0 S(5) lines, we have compared the flux in these lines from the entire nebula,
using the total fluxes derived in \S\ 4.1.2. Because of the very different origin of the 1-0 S(1) to 0-0 S(5) flux ratio, we have not used it in this population distribution
analysis, but it is of interest that it appears that the \twomic\  emission is coming from a region of very similar temperature as the other \htwo\ emission. 

The ratio of intensity of an H$_2$ line produced by a transition from an
upper state $u$ and lower state $l$ is given by
\[
I_{ul} = N (v_u ,J_u)  \, A_{ul} \, h\nu_{ul}
\]
Any intensity can be converted into a column density from this
equation.  Most often we deal with a ratio of intensities, and so
represent the populations as an excitation temperature $T_{u,1;u,2}$,
which is defined as the temperature that produced the derived
population ratio, or implicitly as
\begin{multline*}
\frac{I_{ul, 1}}{I_{ul, 2}} = 
\frac{N_{u,1} \, A_{ul,1} \, h\nu_{ul,1}} {N_{u,2} \, A_{ul,2} \, h\nu_{ul,2}}
\\ = 
\frac{g_{u,1} \, A_{ul,1} \, h\nu_{ul,1}} {g_{u,2} \, A_{ul,2} \, h\nu_{ul,2}} 
 \exp\left[
   \frac{-\chi(v_{u,1} ,J_{u,1}) + \chi(v_{u,2} ,J_{u,2})}{T_{u,1;u,2}}
\right] 
\end{multline*}
where $\chi$ is the excitation energy (K). Our sources of molecular
data for H$_2$ are given in Shaw et~al. (2005).  We use excitation energies given by Dabrowski (1984, corrected
by E. Roueff 2004, private communication). Transition probabilities are taken from Wolniewicz et~al. (1998)
Table 1 gives intensities relative to the 0-0 S(5) line, which has an
upper level of $v=0, J=7$.  Figure 9 shows the derived column densities
relative to the column density of the $v=0,J=7$ level, as a function of
the excitation energy $\chi$.  The lower half of the figure shows the
ratio of column density expressed as an excitation temperature, again
using the previous equation. We do not consider in the solutions for the excitation temperature  the two lines that are each only marginally
detected at a single position or the \twomic\ point derived for the entire nebula. The mean excitation temperatures and standard deviations are also presented in the lower rows of Table 1.  The mean and standard deviation over all
the positions is given as the last pair of rows in the table.

\begin{figure}
  \includegraphics{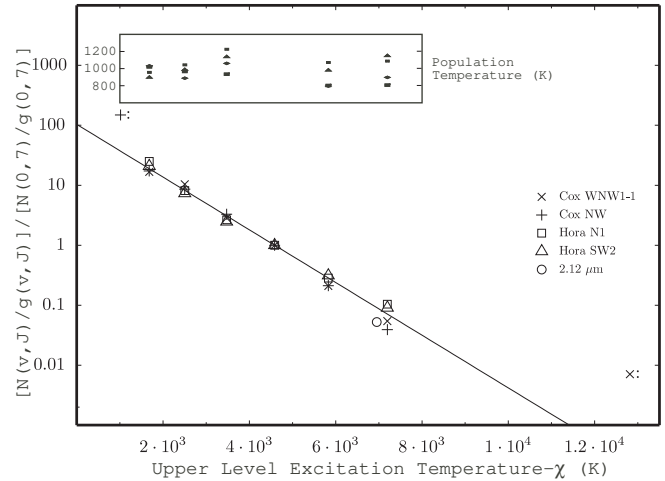}
  \caption{The larger figure shows the population divided by the
    statistical weight in various $v,J$ states relative to that in the
    $v$=0,$J$=7 level (the origin of the 0-0 S(5) transition) as a
    function of the energy of the level expressed in units of the
    excitation temperature, as described in the text.  The line shows
    the relation if all of the levels were characterized by a gas
    temperature of 988 K. The insert gives the gas temperature derived
    for each level, using the 0-0 S(5) line for reference and
    demonstrates the relatively small dispersion around the average
    temperature of 988 K. The +: indicates the position of the 0-0 S(1) line and the 
    o the position of the \twomic\ line, neither of which were used in deriving
    the best fit.}
\end{figure}

\begin{table*}
  \centering
  \caption{Relative Line Fluxes and Derived Temperatures for the \htwo}
  \setlength\tabcolsep{2\tabcolsep}
  \begin{tabular}{ccccccc}\toprule
& & \multicolumn{5}{c}{Relative Flux in Different Samples}\\
& &  Cox & Cox &  Hora &   Hora & Full\\ 
Transition &  $\chi $~(K) &  WNW 1-1 S &   NW    &  N &   SW & Nebula\\ \midrule
0-0 S(1) & 1015 &\nodata & \nodata &  0.229: &\nodata & \\
0-0 S(2) & 1681 &   0.088 &  0.091 &  0.132 &  0.11  & \\
0-0 S(3) & 2503 &   0.916 &  0.757 &  0.716 &  0.645 & \\
0-0 S(4) & 3474 &   0.319 &  0.368 &  0.296 &  0.277 & \\
0-0 S(5) & 4586 &   1     &  1     &  1     &  1     & \\
0-0 S(6) & 5829 &   0.174 &  0.176 &  0.235 &  0.26  & \\
1-0 S(1) & 6951 &   \nodata  &  \nodata  &  \nodata  &  \nodata  & 0.47 \\
0-0 S(7) & 7196 &   0.294 &  0.212 &  0.563 &  0.49  & \\
1-1 S(7) & 12817&   0.033: &\nodata &\nodata &\nodata & \\ \midrule
& $ \rm Derived ~T_\mathrm{excit}$  &
                    935   &  905    & 1040   & 1080 & \\
&        standard deviation     &  110   &  98     & 109    & 97   & \\ \cmidrule{3-6}
& $\langle \rm Derived~T_\mathrm{excit}\rangle$  & \multicolumn{4}{c}{988} & \\
& standard deviation & \multicolumn{4}{c}{119} & \\ \bottomrule
  \end{tabular}
  \label{tab:gary}
\end{table*}

\subsection{Does the weakness of the \htwo\ 1-1 S(7) transition indicate that \htwo\  emission is powered by shocks?}

H06 argues that the weakness of the 1-1 S(7) line shows that the H$_2$ emission must be shock excited 
rather than photo excited. Since their subsequent interpretation of the observed features 
of the nebula are based on this assumption, it merits critical examination. 
H06 state, without presenting detailed proof, that the 1-1 S(7) line is ``normally strong" 
under photo excitation. They did not detect this line, but they did detect the nearby 0-0 S(7) line. Therefore, we base our discussion on the flux ratio of
1-1 S(7) to the 0-0 S(7) line. Examination of their spectra (H06 Figure 9) indicates that the flux ratio must be less than about 0.1. 
The 1-1 S(7) appears to be present in the Cox98 spectra at a level giving a line ratio of about 0.1.

Figure 10 shows the results for a series of calculations in which isothermal clouds
with a range of temperatures and densities was exposed to the radiation field of the central star.
The central star was approximated as a blackbody and the X-ray continua as a series of free-free emitters with the
published luminosities and temperatures.  The total continuum was attenuated by an effective column density of
10$^{22}$ cm$^{-2}$ to approximate the extinction of the ionizing radiation by the H$^{+}$ region.
Typical planetary nebula abundances and ISM grains were assumed.  The clouds had a thickness of $3 \times 10^{15}$ cm. 
The figure shows the 1-1 S(7)/0-0 S(7) line ratio.  The predicted ratio is generally quite small for the values of the \htwo\
density and temperature considered, approaching the upper limit that we identify above only at combinations
of low \htwo\ density and temperature.

\begin{figure}
  \includegraphics{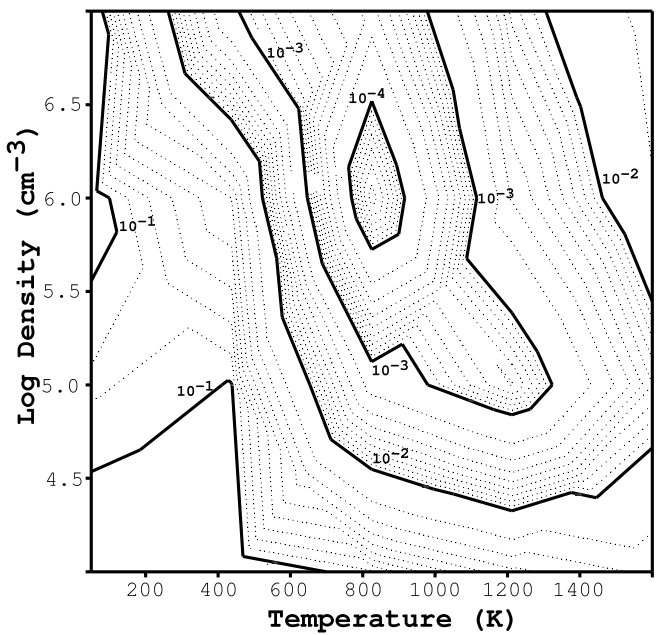}
  \caption{This plot shows the expected flux ratio of the 1-1 S(7) and
    0-0 S(7) lines as a function of the local temperature and
    density in the PDR. }
\end{figure}

Cox98 argue that the total density must exceed 10$^{5}$ \cmq\  in order for collisions to dominate and produce the 
closely single-temperature population distribution. This would be the density to combine with the derived temperature of
900 K for comparison with the S(7) line ratio, if the emitting zone were dominantly molecular hydrogen. In that case the predicted
line ratio is much lower than the upper limit of the observations and one concludes that the observed weakness of the 0-0 S(7) line
is not a useful indicator of the excitation mechanism as assumed by H06.

OHF05 derived a density of the molecular hydrogen from observations of the \twomic\ line, finding a value of $6 \times 10^{4}$ \cmq.
However, they assume that the population of the level producing the transition was in LTE at 900 K,  which seems to be the case
as the upper state producing the \twomic\ line falls right on the population distribution for this temperature.
Using this density would also still indicate that the S(7) line ratio is not a useful
indicator of the excitation mechanism. However, the challenge remains of explaining the unexpected combination of temperature and density. 

Most of the existing calculations of H$_2$ population distributions and resulting emission spectra
have been done for the case of a PDR near an HII region (Black \&\ van Dishoeck 1987). 
The stellar radiation field of an O or B star peaks near the wavelengths that excite the electronic 
transitions of \htwo, about 1000 \AA. The main effect of illumination by this continuum is absorption
into excited electronic states which then decay into excited vibration-rotation levels within the
ground electronic state of \htwo. A very non-thermal distribution is  produced by these electronic photo excitations,
as H06 points out.

However, the Cox98 and H06 spectra discussed in the preceding section
show that the population distribution
within lower levels of \htwo\ is well matched by a thermal distribution at about 900K.  
The non-thermal population distribution produced by O-star photo-excitation is simply not seen. 
This is not surprising since the environment is so dissimilar from that near a main sequence O or B star.
The stellar radiation field of the Helix Nebula central star peaks at much shorter wavelengths than that of an O star,
and \S\ 4.1.2 shows that the observed stellar continuum in the 912-1100  \AA\ interval is too small to account for
the luminosity of the \htwo\ lines by photo-excitation, since the excitation of the fluorescent lines uses
but a small fraction of the total energy in the 912-1100 \AA\ interval.  We agree with H06, that the pure rotational \htwo\
lines are not photon pumped, but not for the reasons they give.
They are far too bright to be produced by photo-excitation by the  current stellar continuum.
We established in \S\  4 that the PDR's of the knots  almost certainly cannot be powered by shocks.
Another energy source is needed.

\subsection{The Nature of the Dissociation Front in the Knots}

In \S\ 4.1.1 and \S\ 4.1.2 we examined the energy budget for the Helix Nebula, establishing that only
the central star radiation more energetic than 13.6 eV  (the extreme ultraviolet radiation or EUV) has enough power to drive the large total flux
of \htwo\ emission that is observed, thus expanding upon and quantifying the conclusions in Cox98. This means that the soft X-ray heating processes and 912-1100 \AA\  (FUV) photo-excitation mechanisms (Natta \&\ Hollenbach 1998) do not explain the Helix observations. The absence of a stellar wind and other time-scale considerations
have already established (OHF05) that shocks cannot be powering the \htwo\ emission and in 
\S\  4.3 we showed why the justification of H06 for a shocks interpretation is incorrect. Although Phillips (2006) established a loose correlation between soft X-ray flux and \htwo\ emission, this correlation is probably secondary rather than primary, as there is insufficient X-ray emission to power the \htwo\ emission; but, the stars that are strong \htwo\ emitters have high temperatures, like the central star in the Helix Nebula. This means that these stars share the property of the EUV radiation being dominant over FUV radiation. Clearly a new
process is required. Based on this observational foundation, we identify a new state of equilibrium that may be common, but has not previously been identified. A new mechanism utilizing the EUV radiation is briefly 
described here and will be elaborated upon in a future  publication.

The ionized flows from the knots are \emph{advection
dominated}, meaning that recombinations are relatively unimportant
(Henney 2001). As a result, neither the FUV  or  X-ray models (Hollenbach  \& Tielens 1997, Natta \&\ Hollenbach 1998) is relevant to the dissociation fronts in the Helix
knots. Instead, the dissociation front merges with the ionization
front (Bertoldi \&\ Draine 1996) and the dissociation of H$_2$ in
this merged front is controlled by the \emph{ionizing} EUV.
The fact that neutral hydrogen 21 cm is not observed in the inner region of the Helix Nebula 
(Rodr\'{\i}guez et~al. 2002) where the optically bright knots are found supports this model and the appearance of
21 cm emission from the more distant and fainter outer-ring knots indicates that a neutral hydrogen
zone is only present there. 
 
The low ionization parameter found in the Helix knots leads to
substantial deviations from ionization and thermal equilibrium since
the dynamical time is shorter than the ionization and heating times.
The effects of this upon the emission from the ionized gas was
discussed at length in OHF05. The dissociation
of H$_2$ in such a front is predominantly due to chemical reactions
with ionized species such as O$^+$, and is therefore a strong function
of the ionization fraction, which is determined by the absorption of
EUV radiation. The radiation field is largely determined by the opacity in the fraction of hydrogen that
is neutral, the key element being its determination of the amount of O$^+$. It is the reaction of O$^+$
with \htwo\ that destroys the \htwo\, rather than the much slower rate of photo-dissociation of \htwo.
This is essentially a one-way process, with \htwo\ entering the zone from the cold molecular core and 
being converted directly to H$^{+}$.
This transition zone is heated by the photo-ionization of neutral hydrogen and can be quite broad and
the preliminary models indicate that it can produce warm \htwo\  column densities of about 10$^{19}$ cm$^{-2}$, as required by the observations (OHF05).  
To the best of our knowledge, no models of such EUV-dominated
dissociation regions have been calculated. We are calculating detailed
models of such regions, which will be reported upon in a future paper and
restrict ourselves here to this brief description.

It is likely that the same process determines the emission seen from the sheath of the tails in \twomic.
The first-order theory for shadowed columns behind optically thick knots was presented by Cant\'o
et~al. (1998). They illustrated that the shadowed regions are illuminated by LyC photons emitted
from recombining hydrogen, that this radiation was closer to 13.6 eV than the ionizing stellar radiation,
and that the flux density of these diffuse LyC photons was about 0.15 that of the direct LyC flux from the 
central star.  These diffuse LyC photons are almost certainly the source of the heating of the \htwo\ in the
tails as the surface brightness in \twomic\ is about 0.23$\pm$0.08 that in the directly illuminated bright cusp (\S\ 3.6).
The implausibility of shocks is also true here and the shortfall of energy from the FUV radiation is even
greater than in the cusps because having a strong diffuse FUV radiation field would demand a large
optical depth in dust for the nebula as a whole, which is not indicated by its emission line spectrum. A separate detailed \htwo\ model is required for the tail because the illuminating FUV will be of lower 
energy and the density much lower than in the bright cusp.

\subsection{The Absence of Radial Features in the Center of the Helix Nebula}

In a recent paper, Meaburn et~al.\@ (2005) presented the analysis of images made in the center of the Helix
Nebula in \hanii\ in 1992 (technical details described in Meaburn et~al. (1998). They present a high contrast
rendering of the image (their Figure 10) and argue that radial ``spokes'' can be seen to faintly continue
inside the boundary of the ``cometary globules'' to within about 30\arcsec\ from the central star.

An arguably superior image of the region is available in the Cerro Tololo Interamerican Observatory 4-m MOSAIC images made in a similar filter
(\Halpha +[NII]) and resolution, with a pixel scale (0.26\arcsec /pixel). The individual exposures were 300 s and in the 
central region, where the fields of the four different pointings overlap, the effective exposure was 1200 s.
A straightforward examination of this image using various levels of brightness and contrast did not reveal the
features posited by Meaburn et~al. (O'Dell 2005). We have now more intensively examined the same images, by
employing median filters of 20x20 pixels and 40x40 pixels and dividing the original images by these, a
technique previously employed (OMM04) to enhance the visibility of radial features in the outer parts of the 
nebula. The results are shown in Figure 11 in negative depiction or the highly 
stretch range of intensities of 0.97-1.03. One sees no indication of radial features extending inside the 
radius at which the knots disappear. We argue that our images are a better test of such features since one
can see numerous stars and galaxies that do not appear in the Meaburn et~al. (2005) image.  

This non-detection of such features, even only slightly inside the position of the easily visible bright cusp
knots is a strong argument that this boundary indicates where knots were first formed and does not represent a 
boundary where knots have been destroyed. An attractive model for generating the initial irregularities that 
develop into the knots is presented in the calculations of Garc\'{\i}a-Segura et~al. (2006), who argue that these 
should arise at the boundary of shocked material as the initial fast-wind phase of the nebula ends,
which is likely to be at about this position within the nebula.

\begin{figure*}
  \includegraphics{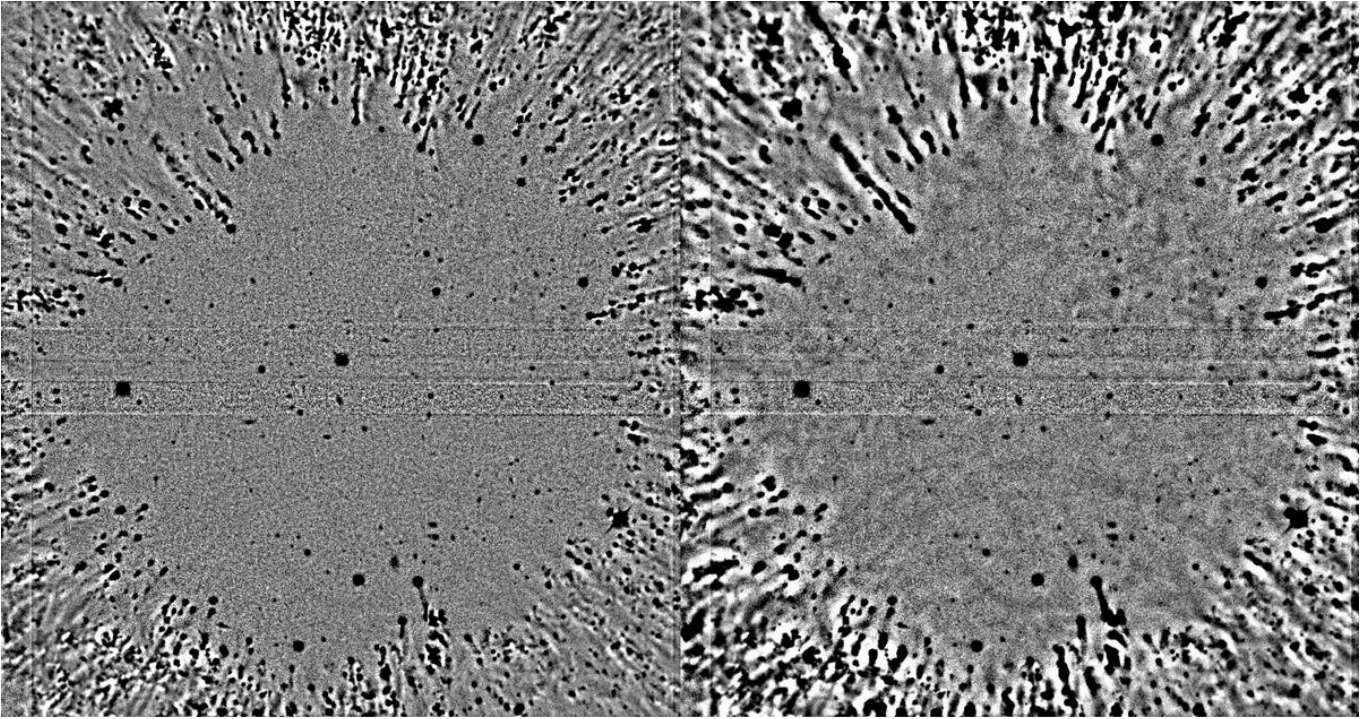}
  \caption{This 377\arcsec x398\arcsec\ \hanii\ image of the center of
    the Helix Nebula shows the ratio of the original image divided by
    a median filtered image of 20x20 pixels (left) and of 40x40 pixels
    (right). No indication of the inner region radial spoke features
    reported by Meaburn et~al. (2005) are seen. The horizontal and
    vertical linear features are the seams of the mosaic formed from
    several CCD's.}
\end{figure*}

\subsection{A Critique of Stream-Source Models for the Helix Knots}

In a series of papers, Dyson and collaborators have developed a model
for the structure of the Helix knots based on the hydrodynamic
interaction that results when ionized gas is injected into a subsonic
stream that flows past the injection source (Dyson et al.\@ 1993, 2006).

Although the injected ionized gas is assumed to arise from an
ionization front on the head of a neutral globule, the radiation
transfer and ionization process is not explicitly included in the
models. This makes it very difficult to make meaningful comparison
between the results of these models and observations of the Helix
nebula. 

Our new \htwo\  observations clearly show (\S\ 3.2)
that the limb-brightened edges of the knot tails correspond to an
ionization front. Additionally, the width of the tail is equal to the
width of the bright cusp at the head of the knot and its conic projection (O'Dell 2000). This would seem to
conclusively establish that radiation shadowing, rather than
hydrodynamic interactions, is the \emph{primary} determinant of the
structure of the tails. In the stream-source model, although tail
widths are predicted to be of the same order as the width of the
injection source, there is no reason to expect them to be equal,
unless the parameters are fine-tuned.

\newcommand\D{\discretionary{}{}{}} 
The kinematic arguments given in support of the stream-source model
(Meaburn et al.\@ 2006) also do not stand up to close
scrutiny. They are based on the ground-based echelle spectroscopic
observations of Meaburn et al.\@ (1998), which seem to show an
acceleration of gas along the tail of the knot 378--801. However, a
comparison with the much higher resolution \textit{HST} observations
(e.g., O'Dell et al.\@ 2005, and also the image available at
http://hubblesite.\D{}org/\D{}gallery/\D{}album/\D{}entire\_collection/\D{}pr199613b).
clearly shows that the sample regions used in that observational study
all correspond to independent knots that are merely projected onto
the tail of 378--801. Therefore, the observed variation in velocity
does not represent an acceleration along the tail, but simply a
tendency towards higher velocities in knots that are farther from the
central star, which has already been noted from CO observations
(Young et al.\@ 1999).

\section{Conclusions}

We have been able to use existing and new observations to reach a number of important conclusions
about the knots in the Helix Nebula.

1. There is sufficient energy to power the nebula's \htwo\ emission only in extreme ultraviolet radiation from the central
star with energies $\ge$13.6 eV, thus eliminating photo-excitation by the 912-1100 \AA\ and X-ray
flux that has been assumed in previous general models.

2. There is no evidence from infrared emission lines for shock excitation of the knots'  \htwo\ emission,
the lack of a driving stellar wind and previous arguments of time-scale having come to the same conclusion.

3. Spectrophotometry of multiple lines in four sample regions and the total nebular flux ratio in the \htwo\ 
0-0 S(5) and \twomic\ lines indicates that the \htwo\ emitting zones are all about 988$\pm$119 K, closely resembling LTE.

4. The \twomic\ emission from individual knots falls immediately inside the ionized gas zone traced by \Halpha\ emission. This is true for both the bright cusps and their fainter tails, the latter establishing that the tails are primarily ionizing radiation shadows, rather than the result of purely hydrodynamic processes.

5. The advection dominated nature of the knot cusps means that there is no
extended neutral hydrogen zone between the cold molecular knot core and the ionized gas layer. This zone of irradiation of \htwo\ by EUV photons is probably the region producing the observed hot gas in the cusps on the star-facing side of the molecular knots and the shadowed regions of the tails.

6. No evidence was found for knots within the He~II core nor were earlier claims verified of linear features extending nearly in to the central star, arguing that the knots have only been created outside the high
ionization core.

\acknowledgments
Anton Koekemoer of the Space Telescope Science Institute provided valuable guidance in the use of the drizzle package of tasks,
making possible the smooth combination of our multiple observations of the same fields.
We are grateful to Angela Speck for providing the calibrated \twomic\ image from her Speck et~al. (2002) paper and to Pierre Cox for his comments on the \htwo\ physical processes.
CRO thanks the Centro de Radioastronom\'\i{}a y Astrof\'\i{}sica,
UNAM, Mexico for generously supporting a two-week visit in February
2006, during which initial work for this paper was carried out and to grant GO 10628 from the Space Telescope Science Institute.
GJF's work was supported in part by grant AR-10653 from the Space Telescope Science Institute,
NASA grant NNG05GD81G, and NSF grant AST 0607028.
WJH acknowledges financial support from DGAPA-UNAM, Mexico, project IN112006.

\end{document}